\documentstyle[12pt]{article}

\makeatletter
\def\eqnarray{\stepcounter{equation}\let\@currentlabel=\theequation
\global\@eqnswtrue
\global\@eqcnt\z@\tabskip\@centering\let\\=\@eqncr
$$\halign to \displaywidth\bgroup\@eqnsel\hskip\@centering
  $\displaystyle\tabskip\z@{##}$&\global\@eqcnt\@ne 
  \hfil$\displaystyle{\hbox{}##\hbox{}}$\hfil
  &\global\@eqcnt\tw@ $\displaystyle\tabskip\z@
  {##}$\hfil\tabskip\@centering&\llap{##}\tabskip\z@\cr}
\def\@sect#1#2#3#4#5#6[#7]#8{\ifnum #2>\c@secnumdepth
    \def\@svsec{}\else
    \refstepcounter{#1}\edef\@svsec{\csname the#1\endcsname.\hskip 1em }\fi
    \@tempskipa #5\relax
    \ifdim \@tempskipa>\z@
    \begingroup #6\relax
    \@hangfrom{\hskip #3\relax\@svsec}{\interlinepenalty \@M #8\par}
    \endgroup
    \csname #1mark\endcsname{#7}\addcontentsline
    {toc}{#1}{\ifnum #2>\c@secnumdepth \else
     \protect\numberline{\csname the#1\endcsname}\fi
           #7}\else
    \def\@svsechd{#6\hskip #3\@svsec #8\csname #1mark\endcsname
          {#7}\addcontentsline
          {toc}{#1}{\ifnum #2>\c@secnumdepth \else
     \protect\numberline{\csname the#1\endcsname}\fi
           #7}}\fi
     \@xsect{#5}}
\def\label#1{\@bsphack\if@filesw {\let\thepage\relax
   \xdef\@gtempa{\write\@auxout{\string
   \newlabel{#1}{{\thesection.\@currentlabel}{\thepage}}}}}\@gtempa
   \if@nobreak \ifvmode\nobreak\fi\fi\fi\@esphack}
\def\@eqnnum{(\thesection.\theequation)}
\def\section{\setcounter{equation}{0} \@startsection {section}{1}{\z@}{-3.5ex
   plus -1ex minus -.2ex}{2.3ex plus .2ex}{\Large\bf}}
\newcount\@minsofar
\newcount\@min
\newcount\@cite@temp
\def\@citex[#1]#2{%
\if@filesw \immediate \write \@auxout {\string \citation {#2}}\fi
\@tempcntb\m@ne \let\@h@ld\relax \def\@citea{}%
\@min\m@ne%
\@cite{%
  \@for \@citeb:=#2\do {\@ifundefined {b@\@citeb}%
    {\@h@ld\@citea\@tempcntb\m@ne{\bf ?}%
    \@warning {Citation `\@citeb ' on page \thepage \space undefined}}%
{\@minsofar\z@ \@for \@scan@cites:=#2\do {%
  \@ifundefined{b@\@scan@cites}%
    {\@cite@temp\m@ne}
    {\@cite@temp\number\csname b@\@scan@cites \endcsname \relax}%
\ifnum\@cite@temp > \@min
    \ifnum\@minsofar = \z@
      \@minsofar\number\@cite@temp
      \edef\@scan@copy{\@scan@cites}\else
    \ifnum\@cite@temp < \@minsofar
      \@minsofar\number\@cite@temp
      \edef\@scan@copy{\@scan@cites}\fi\fi\fi}\@tempcnta\@min
  \ifnum\@minsofar > \z@ 
    \advance\@tempcnta\@ne
    \@min\@minsofar
    \ifnum\@tempcnta=\@minsofar 
      \ifx\@h@ld\relax
        \edef \@h@ld{\@citea\csname b@\@scan@copy\endcsname}%
    \else \edef\@h@ld{\ifmmode{-}\else--\fi\csname b@\@scan@copy\endcsname}%
      \fi
    \else \@h@ld\@citea\csname b@\@scan@copy\endcsname
          \let\@h@ld\relax
  \fi 
\fi}%
\def\@citea{,\penalty\@highpenalty\,}}\@h@ld}{#1}}
\def\appendixname{Appendix}
\def\appendix{\par
  \def\pre@section{\appendixname{}}
  \setcounter{section}{1}
  \@addtoreset{equation}{section}
  \def\thesection{\Alph{section}}
  \def\theequation{\arabic{equation}}}

\def\appendix{\par
  \def\pre@section{\appendixname{}}
  \setcounter{section}{1}
  \@addtoreset{equation}{section}
  \def\thesection{\Alph{section}}
  \def\theequation{\arabic{equation}}}

\makeatother

\begin{document}
\addtolength{\unitlength}{-0.5\unitlength}

\def\b{\beta}
\def\d{\delta}
\def\g{\gamma}
\def\a{\alpha}
\def\s{\sigma}
\def\t{\tau}
\def\l{\lambda}
\def\e{\epsilon}
\def\r{\rho}
\def\d{\delta}
\def\wid{\widehat}
\def\ds{\displaystyle}
\def\be{\begin{equation}}
\def\ee{\end{equation}}
\def\beq{\begin{eqnarray}}
\def\eeq{\end{eqnarray}}
\def\ov{\overline}
\def\om{\omega}

\begin{flushright}
BONN-TH-99-07
\end{flushright}

\rightline{April, 1999}


\phantom{a}

\vspace{2cm}
\centerline{\bf Quantum model of interacting ``strings'' \\
on the square lattice}

\vspace{1cm}

\phantom{a}
\centerline{ H.E. Boos\footnote{
on leave of absence from the Institute for
High Energy Physics, Protvino, 142284, Russia,
E-mail:
boos@avzw02.physik.uni-bonn.de}} 
\centerline{ Physikalisches Institut der Universit{\"a}t Bonn}
\centerline{ 53115, Bonn, Germany }

\phantom{a}

\vspace{0.5cm}

\phantom{a}

\begin{abstract}
{\small
The model which is the generalization of the one-dimensional XY-spin chain for 
the case of the two-dimensional square lattice is considered. The subspace
of the ``string'' states is studied. The solution to the eigenvalue problem
is obtained for the single ``string'' in cases of the ``string'' with fixed 
ends and ``string'' of types (1,1) and (1,2) living on the torus. 
The latter case has
the features of a self-interacting system and looks not to be
integrable while the previous 
two cases are equivalent to the free-fermion model.
}
\end{abstract}

\newpage

\section{Introduction}

In the classical paper \cite{Bethe} Bethe proposed his ansatz
for the eigenfunctions for the one-dimensional Heisenberg chain
\cite{Heischain}.
In the end of this paper he wrote that he intended to generalize
his result for the high dimensional case. Unfortunately,
it was not done. Of course, we do not pretend to solving
this problem. In this paper we just intend to do some steps 
which are more or less in this direction. Namely, we would like to
consider the model proposed by Stroganov \cite{Strog} which can be
thought as a simple generalization of the one-dimensional
spin chain for a two-dimensional case. An idea is to consider
some quantum mechanical model with locally interacting spins.
A Hamiltonian is a sum of some tensor operator products
over the neighbours on a square lattice instead of the 
one-dimensional ordinary spin-chain. Such a Hamiltonian can be considered
as the lattice Hamiltonian for some statistical three-dimensional 
model. For instance, in paper \cite{BQ} Baxter and Quispel obtained the Hamiltonian
for the three-dimensional Zamolodchikov model \cite{Z1,Z2}.

Of course, there are various ways to write down such a Hamiltonian. 
Here we consider only
one of very simple ways to do it. The model is rather similar to the ordinary
free-fermion model \cite{free-ferm} and for some cases we really
reproduce the known answers for the free-fermion model. But inspite
of the simplicity of this model some of cases look not so
trivial and even seem not to be integrable because of some 
self-interacting effects. Nevertheless, we are able to obtain
some equation which we call "secular" equation because of it's
similarity to the ordinary secular equation in quantum mechanics.
In principle, the solution to this equation gives the spectrum
of the model for case of the self-interacting "string"
also.

The paper is organized as follows. In Section 2 we give a formulation
of the model. In Subsection 2.1 we discuss a Hilbert space and
introduce a Hamiltonian. In Subsection 2.2 we give a graphical
interpretation which seems to be useful below. 
In Section 3 we adduce some simple examples.
In Section 4 we consider a diagonalization problem of the Hamiltonian
acting in some closed  subspace of states 
which we call "string" states.  
In Subsection 4.1 we give a basic formulation of "string" states.
In Subsection 4.2 we consider the diagonalization problem of the
"string" Hamiltonian in a case of the "string" with fixed ends.
In Subsection 4.3 we discuss the homology classes $(m,n)$ of the single
"string" living on the torus.
Subsection 4.4 is devoted to the solution of the spectrum problem of
the "string" Hamiltonian in a case when the "string" belongs to the homology
class (1,1). In Subsection 4.5 we consider a case of the "string" from
the homology class (1,2). In Section 5 we discuss the results and possible
ways of a further progress. In Appendix we give the detailed derivation
of the result for the "string" of type (1,2)  
in a case of the $n\times 2$ lattice with cyclic boundary conditions.

\section{Formulation of the model}

\subsection{The quantum state space and the Hamiltonian}

As it was mentioned in Introduction the model \cite{Strog} 
we would like to consider 
is some quantum
mechanical model of two dimensional system of locally interacting
spins. 

Let ${\cal L}$ is simple quadratic $m\times n$ lattice 
with the toroidal topology. Let us enumerate each site of this lattice 
by the pair of integers $(i,j)$ where  
 $0\leq i \leq m-1, 0\leq j \leq n-1$. 
This enumeration is shown in Fig.1 for some part of the lattice 
${\cal L}$.
 
\vspace{0.4cm}
\setlength{\unitlength}{0.35mm}\thinlines
\begin{picture}(200,230)
\multiput(0,180)(8,0){3}{\circle{1}}
\put(25,180){\line(1,0){200}}
\multiput(234,180)(8,0){3}{\circle{1}}
\multiput(0,130)(8,0){3}{\circle{1}}
\put(25,130){\line(1,0){200}}
\multiput(234,130)(8,0){3}{\circle{1}}
\multiput(0,80)(8,0){3}{\circle{1}}
\put(25,80){\line(1,0){200}}
\multiput(234,80)(8,0){3}{\circle{1}}

\multiput(200,30)(0,8){2}{\circle{1}}
\put(200,55){\line(0,1){150}}
\multiput(200,214)(0,8){2}{\circle{1}}
\multiput(150,30)(0,8){2}{\circle{1}}
\put(150,55){\line(0,1){150}}
\multiput(150,214)(0,8){2}{\circle{1}}
\multiput(100,30)(0,8){2}{\circle{1}}
\put(100,55){\line(0,1){150}}
\multiput(100,214)(0,8){2}{\circle{1}}
\multiput(50,30)(0,8){2}{\circle{1}}
\put(50,55){\line(0,1){150}}
\multiput(50,214)(0,8){2}{\circle{1}}

\put(52,182){\tiny$(1,n-1)$}
\put(52,132){\tiny$(0,n-1)$}
\put(52,82){\tiny$(m-1,n-1)$}

\put(102,182){\tiny$(1,0)$}
\put(102,132){\tiny$(0,0)$}
\put(102,82){\tiny$(m-1,0)$}
\put(152,182){\tiny$(1,1)$}
\put(152,132){\tiny$(0,1)$}
\put(152,82){\tiny$(m-1,1)$}
\put(202,182){\tiny$(1,2)$}
\put(202,132){\tiny$(0,2)$}
\put(202,82){\tiny$(m-1,2)$}

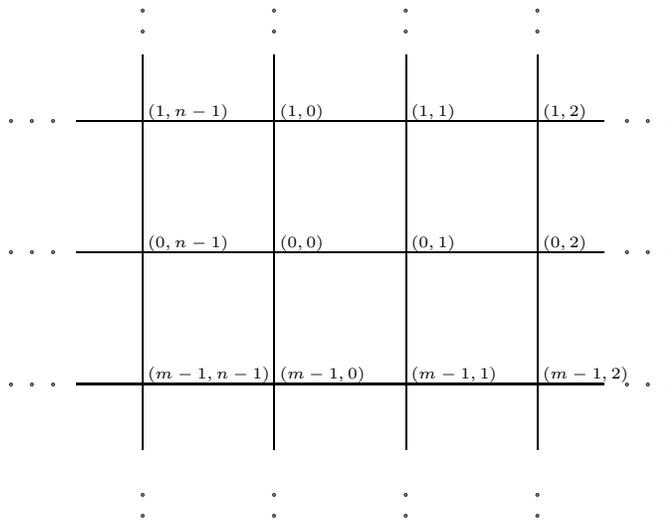
\put(50,10){Fig. 1. Enumeration of the sites.}
\end{picture}

Let us assign some ``spin'' variables with two possible values
(for example, 0 and 1)  to each of   
 $2mn$ edges of the lattice $\cal{L}$.
Let us denote spin variables $\sigma _{i,j}$ for horizontal edges
and $\eta _{i,j}$ for vertical ones as it is shown in Fig. 2.

\vspace{0.4cm}
\setlength{\unitlength}{0.5mm}\thinlines
\begin{picture}(130,130)
\put(30,80){\line(1,0){100}}
\put(80,30){\line(0,1){100}}

\put(30,82){\small$\sigma_{i,j-1}$}
\put(125,82){\small$\sigma_{i,j}$}
\put(82,128){\small$\eta_{i,j}$}
\put(82,30){\small$\eta_{i-1,j}$}

\put(82,82){$(i,j)$}

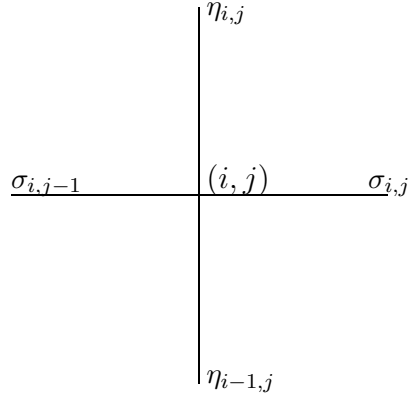
\put(45,10){Fig. 2. Enumeration of the spins.}
\end{picture}

The state space $A$ of our system can be considered as the direct
product of $2mn$ two dimensional complex spaces   
 $G_{i,j}$ and $V_{i,j}$:
$$
A = \prod^{m-1}_{i=0} \prod^{n-1}_{j=0} (\otimes G_{i,j} \otimes V_{i,j}).
$$

Let us choose  $2^{2mn}$ of all possible direct products of the 
basis vectors for the spaces $G_{i,j}$ and $V_{i,j}$ as the basis
vectors for the space $A$.

One can choose the basis vectors in such a way that Pauli matrices have
their standard form:
$$
{\bf \sigma _x} = \left(\begin{array}{cc} 0&1\\
                                          1&0 \end{array} \right),\quad 
{\bf \sigma _y} = \left(\begin{array}{cc} 0&-i\\
                                          i&0 \end{array}  \right),\quad 
{\bf \sigma _z} = \left(\begin{array}{cc} 1&0\\
                                          0&-1  \end{array} \right),\quad 
{\bf \sigma _0} = \left(\begin{array}{cc} 1&0\\
                                          0&1  \end{array} \right). 
$$


The model under consideration is local in a sense that the corresponding
Hamiltonian is decomposed into the sum of the terms which depend on
the fixed number of those spin variables which belong to the 
nearest edges.

Namely, we have
\be
H = \sum^{m-1}_{i=0} \sum^{n-1}_{j=0} h_{i,j}, 
\label{H1}
\ee

where
\be
h_{i,j}=h(\sigma_{i,j};\eta_{i,j};\sigma_{i+1,j};\eta_{i,j+1})
\label{H2}
\ee

The spins taking part in the elementary interaction correspond to the
edges of the elementary sell (square) of the lattice $ {\cal L}$ (see Fig. 3).

\vspace{0.4cm}
\setlength{\unitlength}{0.5mm}\thinlines
\begin{picture}(130,130)
\put(30,50){\line(1,0){90}}
\put(30,100){\line(1,0){90}}
\put(50,30){\line(0,1){90}}
\put(100,30){\line(0,1){90}}

\put(60,105){\small$\sigma_{i+1,j}$}
\put(60,45){\small$\sigma_{i,j}$}
\put(35,75){\small$\eta_{i,j}$}
\put(105,75){\small$\eta_{i,j+1}$}

\put(0,10){Fig. 3. Interacting spins.}
\end{picture}

Now we are ready to fix the exact form of the elementary interaction:

\begin{equation}
\label{ham}
h(\sigma_{(1)},\sigma_{(2)},\sigma_{(3)},\sigma_{(4)}) =
e*(\sigma^-_{(1)}\sigma^+_{(2)}\sigma^+_{(3)}\sigma^-_{(4)}+
\sigma^+_{(1)}\sigma^-_{(2)}\sigma^-_{(3)}\sigma^+_{(4)}),
\label{H3}
\end{equation}

where $\sigma^{\pm}=\sigma^x \pm i\sigma^y$ and the  subscipts in the low
brakets correspond to the four two dimensional spaces:
$G_{i,j}$, $V_{i,j}$, $G_{i+1,j}$ ¨ $V_{i,j+1}$

\subsection{Graphic interpretation}

Let us remind the reader that for the basis of state space $A$ one can 
choose $2^{2mn}$ of the eigenvectors for all spin variables
$\sigma_{i,j}$ and $\eta_{i,j}$. Graphically one can represent each set
of these vectors on the lattice ${\cal L}$ by the colouring all the edges in
one of two possible colours. The colour of each edge is determined by
the eigenvalue of the spin variable which corresponds to this edge.
Some example of such a colouring for a part of the
lattice is shown in Fig.~4.

Instead of two possible colours we shall speak about the presence or
the absence of colour on the corresponding edge of the lattice.
 
\vspace{0.2cm}
\setlength{\unitlength}{0.3mm}\thinlines
\begin{picture}(200,230)
\multiput(0,180)(8,0){3}{\circle{1}}
\put(25,180){\line(1,0){200}}
\multiput(234,180)(8,0){3}{\circle{1}}
\multiput(0,130)(8,0){3}{\circle{1}}
\put(25,130){\line(1,0){200}}
\multiput(234,130)(8,0){3}{\circle{1}}
\multiput(0,80)(8,0){3}{\circle{1}}
\put(25,80){\line(1,0){200}}
\multiput(234,80)(8,0){3}{\circle{1}}

\multiput(200,30)(0,8){3}{\circle{1}}
\put(200,55){\line(0,1){150}}
\multiput(200,214)(0,8){3}{\circle{1}}
\multiput(150,30)(0,8){3}{\circle{1}}
\put(150,55){\line(0,1){150}}
\multiput(150,214)(0,8){3}{\circle{1}}
\multiput(100,30)(0,8){3}{\circle{1}}
\put(100,55){\line(0,1){150}}
\multiput(100,214)(0,8){3}{\circle{1}}
\multiput(50,30)(0,8){3}{\circle{1}}
\put(50,55){\line(0,1){150}}
\multiput(50,214)(0,8){3}{\circle{1}}

\put(102,132){\small$(0,0)$}
\thicklines

\put(99,130){\line(0,1){50}}
\put(149,130){\line(0,1){50}}
\put(150,129){\line(1,0){50}}

\put(50,10){Fig. 4. }
\end{picture}

\vspace{0.5cm}

The basis vector $\mid v> $ which corresponds to Fig.~4 obeys to the
following conditions: 
$$
\begin{array}{ccc}
\sigma_{00} \mid v>=0,\quad &
\sigma_{01} \mid v>=\mid v>,\quad &
\eta_{00} \mid v>=\mid v>,\quad \\

\eta_{01} \mid v>=\mid v>,\quad &
\sigma_{10} \mid v>=0,\quad &
\sigma_{11} \mid v>=0,\quad  .
\end{array}
$$

Let us consider the question of how the Hamiltonian H defined by formulae
(\ref{H1}-\ref{H3}) acts on the
basis vectors. Elementary interaction $h_{i,j}$ depends on those spins
which correspond to the edges of the elementary sell (square) of the
lattice in accordance with the formula \ref{ham} (see Fig. 3).

\vspace{0.2cm}

\setlength{\unitlength}{0.25mm}\thinlines
\begin{picture}(240,230)

\put(30,50){\line(1,0){90}}
\put(150,50){\line(1,0){90}}
\put(30,100){\line(1,0){90}}
\put(150,100){\line(1,0){90}}
\put(30,150){\line(1,0){90}}
\put(150,150){\line(1,0){90}}
\put(30,200){\line(1,0){90}}
\put(150,200){\line(1,0){90}}

\put(50,30){\line(0,1){90}}
\put(50,130){\line(0,1){90}}
\put(100,30){\line(0,1){90}}
\put(100,130){\line(0,1){90}}
\put(170,30){\line(0,1){90}}
\put(170,130){\line(0,1){90}}
\put(220,30){\line(0,1){90}}
\put(220,130){\line(0,1){90}}

\put(30,40){\tiny$(i,j)$}
\put(30,140){\tiny$(i,j)$}
\put(150,40){\tiny$(i,j)$}
\put(150,140){\tiny$(i,j)$}

\put(120,75){\vector(1,0){30}}
\put(120,175){\vector(1,0){30}}

\put(130,80){\tiny$h_{i,j}$}
\put(130,180){\tiny$h_{i,j}$}
\thicklines

\put(50,99){\line(1,0){50}}
\put(50,149){\line(1,0){50}}
\put(170,49){\line(1,0){50}}
\put(170,199){\line(1,0){50}}

\put(49,50){\line(0,1){50}}
\put(99,150){\line(0,1){50}}
\put(219,50){\line(0,1){50}}
\put(169,150){\line(0,1){50}}

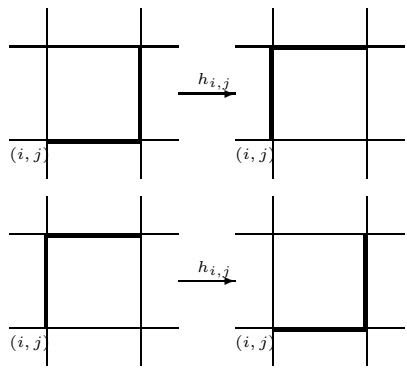
\put(30,10){Fig. 5. The action of the Hamiltonian $h_{i,j}$.}
\end{picture}

The operators $\sigma ^+$ ``give the colour'' to the corresponding edge
without colour 
and annihilate the state which corresponds to the coloured edge. In contrast,
the operators $\sigma ^-$ ``remove the colour'' from the coloured edge
and annihilate the state which corresponds to the edge without colour. 
The action of the $h_{i,j}$ on the colouration of the elementary square
(sell) can be illustrated by Fig. 5.

\section{Some simple examples}

1. There are a lot of quantum states corresponding to some special
colouring of the lattice which are ``unmovable'', i.e. the Hamiltonian
acts on such a states trivially just annihilating them.
The simplest example of such a state is a straight line or the set of
straight lines not intersecting with each other or intersecting in the
end points as it is shown in Fig. 6

\begin{picture}(400,250)
\multiput(60,60)(20,0){16}{\line(0,1){180}}
\multiput(60,60)(0,20){10}{\line(1,0){300}}
\thicklines
\multiput(140,80)(1,0){2}{\line(0,1){120}}
\multiput(160,120)(1,0){2}{\line(0,1){100}}
\multiput(200,160)(0,1){2}{\line(1,0){60}}
\multiput(200,160)(1,0){2}{\line(0,1){40}}
\multiput(240,100)(0,1){2}{\line(1,0){100}}
\put(180,20){Fig. 6}
\end{picture}

Some other examples of ``unmovable'' states are shown in Fig. 7

\setlength{\unitlength}{0.23mm}\thinlines

\begin{picture}(400,215)
\multiput(100,40)(20,0){14}{\line(0,1){160}}
\multiput(100,40)(0,20){9}{\line(1,0){260}}
\thicklines
\multiput(120,80)(1,0){2}{\line(0,1){20}}
\multiput(120,80)(0,1){2}{\line(1,0){60}}
\multiput(120,100)(0,1){2}{\line(1,0){40}}
\multiput(180,80)(1,0){2}{\line(0,1){60}}
\multiput(160,100)(1,0){2}{\line(0,1){20}}
\multiput(140,120)(1,0){2}{\line(0,1){20}}
\multiput(140,120)(0,1){2}{\line(1,0){20}}
\multiput(140,140)(0,1){2}{\line(1,0){40}}

\multiput(120,160)(1,0){2}{\line(0,1){20}}
\multiput(120,160)(0,1){2}{\line(1,0){20}}
\multiput(120,180)(0,1){2}{\line(1,0){20}}
\multiput(140,160)(1,0){2}{\line(0,1){20}}

\multiput(160,160)(0,1){2}{\line(1,0){60}}
\multiput(160,160)(1,0){2}{\line(0,1){20}}
\multiput(160,180)(0,1){2}{\line(1,0){60}}
\multiput(220,160)(1,0){2}{\line(0,1){20}}

\multiput(260,160)(1,0){2}{\line(0,1){20}}
\multiput(260,160)(0,1){2}{\line(1,0){60}}
\multiput(260,180)(0,1){2}{\line(1,0){80}}
\multiput(320,140)(0,1){2}{\line(1,0){20}}
\multiput(320,140)(1,0){2}{\line(0,1){20}}
\multiput(340,140)(1,0){2}{\line(0,1){40}}

\multiput(220,120)(0,1){2}{\line(1,0){40}}
\multiput(220,120)(1,0){2}{\line(0,1){20}}
\multiput(220,140)(0,1){2}{\line(1,0){80}}
\multiput(300,60)(1,0){2}{\line(0,1){80}}

\put(220,0){Fig. 7}
\end{picture}

\vspace{1cm}

2. Now let us adduce some simple examples of quantum states on which
the Hamiltonian	acts non-trivially.
In fact, the simplest case of such a states
is shown in Fig 5. Namely, in this case there are only two possible
quantum states, say $\mid 0>$ and $\mid 1>$, which
correspond respectively to the right and left sides
of the first line in Fig. 5 where only two edges of the whole lattice
are coloured. 
The parameter $e$ which is the single dimensional parameter in our
problem (see (\ref{ham})) can be equated to unity without loss of 
generality.
Then the Hamiltonian $H$ just intergchanges these two states:

\beq
&H\;\mid 0>\;=\;\mid 1>&\nonumber\\
&H\;\mid 1>\;=\;\mid 0>&
\label{f1}
\eeq
It is very easy to find the eigenvalues $E$ of the Hamiltonian in this case
\be
E\;=\;\pm 1.
\label{E0}
\ee

As the next case let us consider ``one-step wave'' which is shown in Fig.~8

\vspace{0.5cm}

\setlength{\unitlength}{0.3mm}\thinlines
\begin{picture}(400,140)
\multiput(60,60)(20,0){13}{\line(0,1){60}}
\multiput(60,60)(0,20){4}{\line(1,0){240}}
\thicklines
\multiput(100,80)(0,1){2}{\line(1,0){60}}
\multiput(160,100)(0,1){2}{\line(1,0){100}}
\multiput(160,80)(1,0){2}{\line(0,1){20}}
\put(100,40){0}
\put(160,40){$\l$}
\put(250,40){$N$}
\put(120,0){Fig.8 "One-step wave"}
\end{picture}

\vspace{1cm}

Let $N$ be the ``distance'' between the end points of this ``wave''
 and $\l$ is the
coordinate of the ``jump'' (in Fig. 8 $\l=3,N=8$).
 Let us denote this state $\mid \l>$.  
It is easy to see that the Hamiltonian acts on this state as follows:
\be
H\mid \l>\;=\;\mid \l+1>\;+\;\mid \l-1>
\label{f2}
\ee

To solve the eigenvalue problem $H\mid \Psi>=E\mid \Psi>$ we can try
to do the following substitution

\be
\mid \Psi>=\sum_{\l=0}^N \Psi(\l)\mid \l>.
\label{basis0}
\ee

The static Schr{\"o}dinger equation for the wave function looks very simple

\begin{equation}
\label{eqgennew0}
E\Psi (\l)=\Psi (\l+1)+\Psi (\l-1).
\end{equation}

Two states  $\l=0$ and $\l=N$ are 
the ``first'' and ``last'' states of the wave.
Hence, we have the following "boundary" conditions for the wave function
\be
\Psi(-1)\;=\;\Psi(N+1)\;=0.
\label{b1}
\ee
The eigenvalues $E$ and eigenfunctions of the Hamiltonian can be easily found
\be
E\;=\;x\,+\,x^{-1}
\label{E1}
\ee
and
\be
\Psi( \l)\;=\;x^{\l+1}-x^{-\l-1}.
\label{V1}
\ee
In order to satisfy the conditions (\ref{b1}) we should set $x$ to be some root of unity
\be
x^{2N+4}\;=\;1
\label{x1}
\ee
or
\be
x\;=\;e^{{{i\pi k}\over{N+2}}}
\label{x1a}
\ee
and $k=1,\ldots,N+1$. Then $E\;=\;2\cos{{{i\pi k}\over{N+2}}}$. 
The previous case corresponds to $N=1$. Hence we have only two 
possibilities $k=1$ and $k=2$. In the first case $E=2\cos{{{i\pi}\over 3}}=1$.
In the second case $E=2\cos{{{2i\pi}\over 3}}=-1$ in accordance with
(\ref{E0}). 

We do not intend to classify here the whole space of  quantum states. Below we
consider only some important subspace of states which we call "string" states.

\section{The string spectrum}

\subsection{``Ice condition'' and ``string'' states}

Let us consider the subspace of the state space which is the linear
span of some special subset of the basis vectors.
Namely, let us consider all basis vectors for which the corresponding
colouring of the lattice ${\cal L}$ satisfies to the so-called ``ice
condition'' (see, for example, \cite{Bax}). In other words, there are
only six types of the allowed colouring of the edges which are 
adjacent with one site of the lattice (six types of the vertices
in accordance with the standard terminology of statistical models
on the squared lattice), see Fig. 9.
 
\vspace{0.2cm}
\setlength{\unitlength}{0.8mm}
\begin{picture}(140,50)
\put(30,1){Fig. 9. Ice condition for the verticies.}
\put(20,30){\circle*{2}}
\put(40,30){\circle*{2}}
\put(60,30){\circle*{2}}
\put(80,30){\circle*{2}}
\put(100,30){\circle*{2}}
\put(120,30){\circle*{2}}

\put(19,18){1}
\put(39,18){2}
\put(59,18){3}
\put(79,18){4}
\put(99,18){5}
\put(119,18){6}

\thicklines
\put(20,29.8){\line(1,0){8}}
\put(20,30.2){\line(1,0){8}}
\thinlines
\put(40,30){\line(1,0){8}}
\put(60,30){\line(1,0){8}}
\thicklines
\put(80,29.8){\line(1,0){8}}
\put(100,29.8){\line(1,0){8}}
\put(80,30.2){\line(1,0){8}}
\put(100,30.2){\line(1,0){8}}
\thinlines
\put(120,30){\line(1,0){8}}
\thicklines
\put(20,29.8){\line(-1,0){8}}
\put(20,30.2){\line(-1,0){8}}
\thinlines
\put(40,30){\line(-1,0){8}}
\put(60,30){\line(-1,0){8}}
\thicklines
\put(80,29.8){\line(-1,0){8}}
\put(80,30.2){\line(-1,0){8}}
\thinlines
\put(100,30){\line(-1,0){8}}
\thicklines
\put(120,29.8){\line(-1,0){8}}
\put(120,30.2){\line(-1,0){8}}
\put(20.2,30){\line(0,1){8}}
\put(19.8,30){\line(0,1){8}}
\thinlines
\put(40,30){\line(0,1){8}}
\thicklines
\put(59.8,30){\line(0,1){8}}
\put(60.2,30){\line(0,1){8}}
\thinlines
\put(80,30){\line(0,1){8}}
\put(100,30){\line(0,1){8}}
\thicklines
\put(119.8,30){\line(0,1){8}}
\put(120.2,30){\line(0,1){8}}
\put(19.8,30){\line(0,-1){8}}
\put(20.2,30){\line(0,-1){8}}
\thinlines
\put(40,30){\line(0,-1){8}}
\thicklines
\put(59.8,30){\line(0,-1){8}}
\put(60.2,30){\line(0,-1){8}}
\thinlines
\put(80,30){\line(0,-1){8}}
\thicklines
\put(99.8,30){\line(0,-1){8}}
\put(100.2,30){\line(0,-1){8}}
\thinlines
\put(120,30){\line(0,-1){8}}
\end{picture}

\vspace{1cm}

The dimension of this subspace ${\cal A}$ is equal to the number of
the ways to colour the lattice ${\cal L}$ taking into account the 
ice condition. This quantity is connected with the entropy
of the ``ice'' model which was calculated in the thermodynamic limit
by Lieb \cite{Lieb}. 

The action of the Hamiltonian $H$ conserves the ice condition.
In other words, the subspace ${\cal A}$ is invariant under the action of $H$.
Our main problem is to find the eigenvalues of the Hamiltonian $H$
on the subspace ${\cal A}$.

The subspace ${\cal A}$ in it's turn is the direct sum of the invariant
subspaces ${\cal A}^{(k)}$ where $k=0,1,2,...$ are nonnegative integers.

Any configuration which satisfies the ice condition
can be represented as a set of continuous nonintersecting ways 
which go only up and to the left (see, for example, \cite{Bax}).
 
Let us call this ways  ``stings''. Also, let us denote
${\cal A}^{(k)}$ the subspaces which are spanned on the
vectors  corresponding to the colouring the lattice by
$k$ ways (or stings). We call ${\cal A}^{(k)}$ the 
state space with $k$ strings.

It is easy to see that all subspaces ${\cal A}^{(k)}$ are invariant
under the action of the Hamiltonian $H$. So, the number of strings
$k$ is the conservation number.

Since the subspace ${\cal A}^{(0)}$ consists of only one vector 
$\mid 0>$ which is annihilated by the Hamiltonian  $H \mid 0>=0$
let us begin the solving our problem of the diagonalization of $H$
from the subspace ${\cal A}^{(1)}$. So, let us consider 
 the spectrum for the single string. 

\subsection{The string with the ``fixed'' ends}

In Section 3 we considered the case of the ``one-step wave''.
In fact, this is a simple example of a string with the fixed ends
when the string has only one ``jump''. 
Suppose that 
the size of the lattice is big enough
to consider more
general case when the string has $N$ ``jumps''. The quantum state
can be defined by fixing the coordinates $\l_1,\l_2,\ldots \l_M$ 
of these jumps. Due to the ``ice'' condition these coordinates should
satisfy the ordering:
\be
0\leq\l_1\leq\l_2\leq\ldots\leq\l_M\leq N
\label{order1}
\ee
which corresponds to some partition $\l=(\l_1,\l_2,\ldots,\l_M)$.
\footnote{see for example the book \cite{Mac}}
Some example of the string state for the case $N=9$ is shown in Fig. 10

\addtolength{\unitlength}{-0.6\unitlength}

\begin{picture}(400,250)
\multiput(100,80)(20,0){12}{\line(0,1){160}}
\multiput(100,80)(0,20){9}{\line(1,0){220}}
\thicklines
\multiput(120,100)(0,1){2}{\line(1,0){20}}
\multiput(140,100)(1,0){2}{\line(0,1){40}}
\multiput(140,140)(0,1){2}{\line(1,0){20}}
\multiput(160,140)(1,0){2}{\line(0,1){20}}
\multiput(160,160)(0,1){2}{\line(1,0){40}}
\multiput(200,160)(1,0){2}{\line(0,1){20}}
\multiput(200,180)(0,1){2}{\line(1,0){40}}
\multiput(240,180)(1,0){2}{\line(0,1){40}}
\multiput(240,220)(0,1){2}{\line(1,0){60}}
\put(115,60){0}
\put(130,60){$\l_1$}
\put(130,40){$\l_2$}
\put(160,60){$\l_3$}
\put(200,60){$\l_4$}
\put(230,60){$\l_5$}
\put(230,40){$\l_6$}
\put(285,60){$N$}
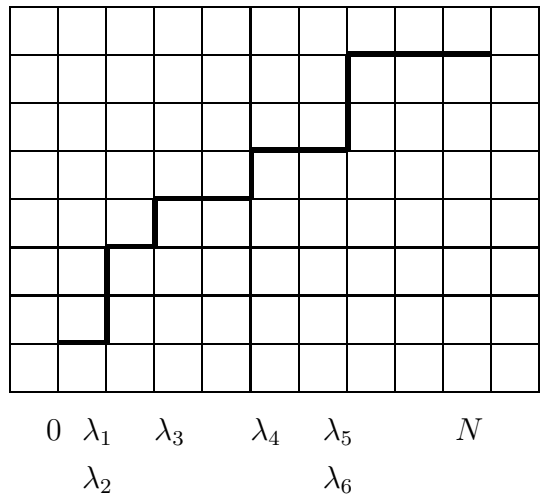
\put(30,20){Fig.10 The string corresponding to the partition $(1,1,2,4,6,6)$}
\put(100,0){in the case $N=9$}
\end{picture}

\vspace{1cm}

Let us fix $N$ and $M$.
It is not difficult to calculate the dimension of the quantum space of the single
string with the fixed ends. To do it let us consider $\ov\l_j=\l_j+j$ which
satisfy the chain of the strict inequalities instead of the non-strict ones
given by (\ref{order1})
\be
1<\ov\l_1<\ov\l_2<\ldots<\ov\l_M< N+M+1.
\label{order2}
\ee
So, the problem is reduced to the calculating  the number of the different
ways to put $M$ undistinguishable objects on the $N+M$ places.
The answer is well known. It is the binomial coefficient $N+M\choose M$.
Therefore we have for a dimension of the quantum space
\be
dim\; = \; {N+M\choose M}.
\label{dim1}
\ee

It is easy to see that the Hamiltonian acts on some arbitrary string state $\l$
as follows

\be
H\,\mid \l>\;=\;\sum_{i=1}^M \;\; \mid \l+\d_i>\;+\;\mid \l-\d_i>
\label{f3}
\ee
where $\d_i=(0,\ldots,0,1,0,\ldots,0)$ is 
$M$-component vector which has $1$ on the $i^{-th}$ place and zeros
on other places.

Now let us consider the wave functions instead of the basis
vectors. As in Section 3 we are going to find the eigenvectors for the Hamiltonian\\
$H\mid \Psi>=E\mid \Psi>$ in the following form:

\be
\mid \Psi>=\sum_{\l_1,...\l_M}\Psi(\l)\mid \l>
\label{basis}
\ee
where $\Psi(\l)$ is short notation for the wave function $\Psi(\l_1,\ldots,\l_N)$.
The Schr{\"o}dinger equation in the general position looks as follows:

\begin{equation}
\label{eqgennew}
E\,\Psi (\l)=\sum^M_{i=1}
\Psi (\l+\d_i)+
\Psi (\l-\d_i).
\end{equation}

Of course, the states in the RHS of (\ref{f3}) or (\ref{eqgennew}) should also correspond
to some partition with the ordering (\ref{order1}). 
We can formally satisfy this
requirement by equating  to zero the wave functions which correspond to
non-existing states $\mid \l>$ with $\l_{i+1}=\l_i-1$.
As in the case of the ``one-step wave'' the ends of the string can not ``move''.
We can again formally equate the wave functions with $\l_1=-1$ and $\l_{M}=N+1$ to zero.
So, we have

\beq
\Psi(\l)\;=\;0,\quad \mbox{if} \quad
\cases{ & $\>\l_{i+1}=\l_i-1\>$ for at least one position $i$ \cr
 & $\quad\l_1=-1$ \cr
 & $\quad\l_{M}=N+1$\cr}
\label{b2}
\eeq

It is easy to see that for the string of the general position $0<\l_1<\ldots<\l_M$
the monomial solution
\be
\Psi(\l)\;=\;x^{\l}\;=\;x_1^{\l_1}\ldots x_M^{\l_M}
\label{mon1}
\ee
satisfies the eigenvalue problem for the Hamiltonian (\ref{f3})
with the eigenvalue (or energy)
\be
E\;=\;\sum_{i=1}^M x_i\,+\,x_i^{-1}
\label{E2}
\ee
which is invariant under any permutation of the variables $x_i$
and a substitution $x_i\rightarrow x_i^{-1}$.
Of course, the monomial solution (\ref{mon1}) does not 
satisfy the conditions (\ref{b2}). To find the right solution
to the eigenvalue problem we have to consider a Bethe ansatz-like linear
combination
\be
\sum_{\s,\e_1,\ldots\e_M}A_{\s}(\e_1,\ldots\e_M)x_{\s(1)}^{\e_1\l_1}
\ldots x_{\s(M)}^{\e_M\l_M}
\label{pol1}
\ee
where the sum is  over all permutation $\s$ and all possible
signs $\e_i=\pm 1$.
One can check that the following simple determinant solution
\be
\Psi(\l)\;=\;\mbox{Det}([x_i^{\l_j+j}])_{1\leq i,j\leq M}
\label{det1}
\ee
where $[x]=x-x^{-1}$ 
belongs to the class of polynomials given by (\ref{pol1})
and satisfy almost all conditions (\ref{b2}) with the exception
of the last one in (\ref{b2}). 
Indeed, when we try to disturb the order of the sequence of parameters
$\l_i$ we get the pair of the same coloumns. So, the unwanted moves
are suppressed automaticaly!

Hence, we have to satisfy the conditions $\Psi(\l)=0$ for $\l_M=N+1$.
So, we come to the quantization condition
\be
x_i^{2\,(N+M+1)}\;=\;1
\label{x2}
\ee
which generalizes the condition (\ref{x1}) for the "one-step wave" case.
To enumerate all eigenvectors we should set
\be
x_i\;=\;\Omega^{k_i}
\label{x2a}
\ee
where $\Omega=e^{i\pi/(N+M+1)}$ and the wave numbers $k_i$ satisfy
the inequalities
\be
0<k_1<k_2<\ldots<k_M<N+M+1.
\label{k1}
\ee
This chain of the inequalities coincides with the chain for $\ov\l_j$
given by (\ref{order2}). Hence, the number of possible values $k_i$
i.e. the number of all eigenvectors coincides with the dimension of
the quantum space given by formula (\ref{dim1}) as it should be.

The corresponing values of the energy $E$ are still given by (\ref{E2})
which can be written in the following form
\be
E\;=\;\sum_{j=1}^{M}\,2\,\cos{{\pi k_j}\over{N+M+1}}.
\label{E2a}
\ee
The appearence of the determinant in formula (\ref{det1}) for the
eigenfunctions tells us about the connection of this problem
with the free-fermion model or $XY$-chain \cite{free-ferm}. 

In the next section we consider the
spectral problem for the single string on the lattice with toroidal
boundary conditions.

\subsection{The ring of homologies of the single string on a torus} 

Let us consider the arbitrary single string configuration on the
$m \times n$ lattice torus ${\cal L}$. Let us apply to the 
figure 11 

\vspace{0.4cm}
\setlength{\unitlength}{0.3mm}
\begin{picture}(240,190)
\put(50,1){Fig. 11 }

\thinlines
\put(30,30){\line(1,0){210}}
\put(30,60){\line(1,0){210}}
\put(30,90){\line(1,0){210}}
\put(30,120){\line(1,0){210}}
\put(30,150){\line(1,0){210}}
\put(30,180){\line(1,0){210}}

\put(30,30){\line(0,1){150}}
\put(60,30){\line(0,1){150}}
\put(90,30){\line(0,1){150}}
\put(120,30){\line(0,1){150}}
\put(150,30){\line(0,1){150}}
\put(180,30){\line(0,1){150}}
\put(210,30){\line(0,1){150}}
\put(240,30){\line(0,1){150}}

\thicklines
\put(120,59){\vector(1,0){60}}
\put(30,89){\vector(1,0){150}}
\put(180,89){\vector(1,0){60}}
\put(30,149){\vector(1,0){90}}
\put(180,149){\vector(1,0){60}}

\put(119,30){\vector(0,1){30}}
\put(119,150){\vector(0,1){30}}
\put(179,60){\vector(0,1){30}}
\put(179,90){\vector(0,1){60}}
\end{picture}

\vspace{1cm}

One can think that we see the piecies of the three strings
if to consider this figure as the part of some bigger lattice. But
it turns out to be obvious that it is one string living on the 
$5 \times 7$ torus. This string winds round the torus 
$\overline m=2$ times on the horizontal direction and $\overline n=1$ times
on the vertical direction.

The Hamiltonian being local does not change the winding number
of the string. So, the space ${\cal A}^{(1)}$  decomposes
into the direct sum of the invariant subspaces $a_{\overline m,\overline n}$.

What can the integers $\overline m$ and $\overline n$ be ?
Let $\overline m=0$ then $\overline n=1$ and we have $n$ of the trivial
string configurations of the following form:

\vspace{0.4cm}
\setlength{\unitlength}{0.2mm}
\begin{picture}(500,200)
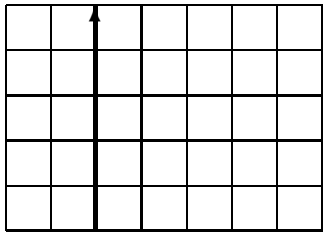
\put(130,-10){  Fig. 12. Simple winding of the type (0,1)}

\thinlines
\put(230,30){\line(1,0){210}}
\put(230,60){\line(1,0){210}}
\put(230,90){\line(1,0){210}}
\put(230,120){\line(1,0){210}}
\put(230,150){\line(1,0){210}}
\put(230,180){\line(1,0){210}}

\put(230,30){\line(0,1){150}}
\put(260,30){\line(0,1){150}}
\put(290,30){\line(0,1){150}}
\put(320,30){\line(0,1){150}}
\put(350,30){\line(0,1){150}}
\put(380,30){\line(0,1){150}}
\put(410,30){\line(0,1){150}}
\put(440,30){\line(0,1){150}}

\thicklines

\put(289,30){\vector(0,1){150}}
\end{picture}

\vspace{1cm}

In the same way we have $\overline m=1$ if $\overline n=0$ and
we have $m$ of the trivial horizontal configurations.

Let us note without the proof the fact that the space
${\cal A}^{(1)}$ decomposes into the direct sum of $mn+2$ invariant
subspaces $a_{0,1}$, $a_{1,0}$ $\ldots$
$a_{\overline m, \overline n}$ where
$\overline m$ and $\overline n$ satisfy the following
inequalities:

$$
1 \leq \overline m \leq m, \quad 1 \leq \overline n \leq n.
$$

In the next section we shall consider the simplest nontrivial case
of the string of the type (1,1). This string winds the torus one time
in the horizontal direction and one time in the vertical one.

We shall see that the string in this case
is also equivalent to the $XY$-chain.
It will be also equivalent to the model considered by Bethe if to add
the diagonal part to the Hamiltonian $H$. 

\subsection{The spectrum of the string of type (1,1)}

So, let us consider the subspace $a_{1,1}$.
One can split all configurations of the string on $m$ of classes
which depend on the number of the horizontal line in the torus where
the string leaves or intersects the vertical line with the number 0. 
The number of the class will be denoted as $\alpha$:

$$
\alpha =0,1,...,m-1.
$$

 Fig. 12 corresponds to the case $\alpha=1$ (for the torus $4 \times 7$).

\vspace{0.3cm}
\setlength{\unitlength}{0.3mm}
\begin{picture}(240,180)
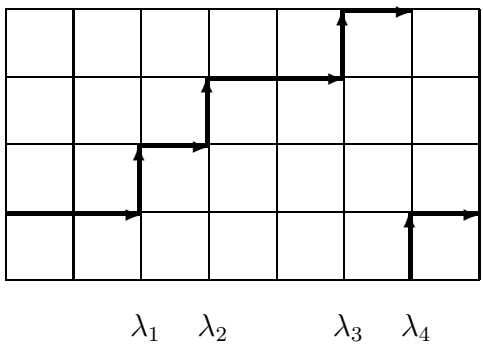
\put(40,-20){  Fig. 13. The marking of the configuration (1,1)}

\thinlines
\put(30,30){\line(1,0){210}}
\put(30,60){\line(1,0){210}}
\put(30,90){\line(1,0){210}}
\put(30,120){\line(1,0){210}}
\put(30,150){\line(1,0){210}}

\put(30,30){\line(0,1){120}}
\put(60,30){\line(0,1){120}}
\put(90,30){\line(0,1){120}}
\put(120,30){\line(0,1){120}}
\put(150,30){\line(0,1){120}}
\put(180,30){\line(0,1){120}}
\put(210,30){\line(0,1){120}}
\put(240,30){\line(0,1){120}}

\thicklines

\put(30,59){\vector(1,0){60}}
\put(90,89){\vector(1,0){30}}
\put(120,119){\vector(1,0){60}}
\put(180,149){\vector(1,0){30}}
\put(210,59){\vector(1,0){30}}

\put(89,60){\vector(0,1){30}}
\put(119,90){\vector(0,1){30}}
\put(179,120){\vector(0,1){30}}
\put(209,30){\vector(0,1){30}}

\put(85,5){$\l_1$}
\put(115,5){$\l_2$}
\put(175,5){$\l_3$}
\put(205,5){$\l_4$}

\end{picture}

\vspace{1cm}

Again the configuration for some fixed $\alpha$ is characterized by $m$ of
the integers:

\begin{equation}
\label{neq1}
1 \leq \l_1 \leq \l_2 \leq ... \leq \l_m \leq n
\end{equation}
where as in the case of the string with fixed end points 
$\l_i$  is the number of those vertical line where the string
begins it's ``climbing up'' on the $i^{-th}$ sell from the horizontal
line with the number  $\alpha$. For example, Fig. 12 corresponds to
the following set $\l_1=2$, $\l_2=3$, $\l_3=5$ and $\l_4=6$ i.e.
to the partition $(2,3,5,6)$.

To calculate the whole number of the single string
configurations we can apply a similar trick as in the case
of string with the fixed ends.
Namely, the chain of the non-strict inequalities (\ref{neq1}) is
equivalent to the following chain of the strict inequalities:

\be
0<\l_1<\l_2+1<\l_3+2<...<\l_m+m-1<n+m
\label{lambda}
\ee

Hence, we can apply the same combinatorial calculation as it was
done for derivation of formula (\ref{dim1}).
Taking into consideration that the variable $\alpha$ takes $m$ 
different values we get the dimension of the quantum space of the string
of type (1,1):

\be
dim(a_{1,1})=\frac{(m+n-1)!}{(n-1)!(m-1)!}.
\label{dim2}
\ee

As it was mentioned above the calculation of all possible configurations with
any number of the strings is equivalent to the problem of the calculation
of the entropy in the ice model on the square lattice.

Now we return to the problem of the diagonalization of the Hamiltonian
 $H$. Let us introduce the following notations for the basis
vectors:

\be
\mid \alpha ;\l>\;=\;\mid \alpha ;\l_1,\l_2,...,\l_m>.
\label{stringstate}
\ee

This vector corresponds to the single string configuration of the type (1,1)
from the class 
$\alpha$ which is caracterized by the parameters $\l_1,\l_2,...\l_m$.

Let us say that the parameters $\l_i$ are on the general position if
none of equalities in the chain of inequalities (\ref{neq1})
is fulfiled. The action of the Hamiltonian on the states
parameterized by the parameters $\l_i$ in the general position is given
by a similar to (\ref{f3}) formula 

\begin{equation}
\label{eqgen}
H\mid \alpha ;\l>=\sum^m_{i=1}
\mid \alpha ;\l+\d_i>+
\mid \alpha ;\l-\d_i>.
\end{equation}

Then suppose there is some group of the coinciding parameters
$\l_j,\l_{j+1},...\l_k$ where $j<k$ but $\l_j>1$ and $\l_k<n$.
This configuration corresponds to the presence of the jump
of the length $k-j+1$ (see Fig. 14).

\vspace{0.3cm}
\setlength{\unitlength}{0.3mm}
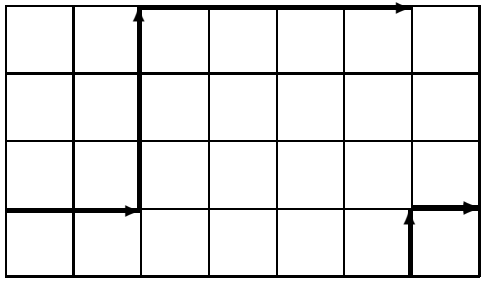
\begin{picture}(240,170)
\put(40,-10){  Fig. 14. The case of coinciding $\l_i$}

\thinlines
\put(30,30){\line(1,0){210}}
\put(30,60){\line(1,0){210}}
\put(30,90){\line(1,0){210}}
\put(30,120){\line(1,0){210}}
\put(30,150){\line(1,0){210}}

\put(30,30){\line(0,1){120}}
\put(60,30){\line(0,1){120}}
\put(90,30){\line(0,1){120}}
\put(120,30){\line(0,1){120}}
\put(150,30){\line(0,1){120}}
\put(180,30){\line(0,1){120}}
\put(210,30){\line(0,1){120}}
\put(240,30){\line(0,1){120}}

\thicklines

\put(30,59){\vector(1,0){60}}
\put(90,149){\vector(1,0){120}}

\put(89,60){\vector(0,1){90}}
\put(209,30){\vector(0,1){30}}

\put(210,61){\vector(1,0){30}}
\put(210,60){\vector(1,0){30}}

\put(65,10){$\l_1=\l_2=\l_3$}
\put(205,10){$\l_4$}

\end{picture}

\vspace{1cm}

For this group of the parameters evidently only
the decrease of $\l_j$ and the increase of $\l_k$
on the unity are allowed. Let us note that after this the parameter $\alpha$ 
is not changed. 

Now let us consider the situation which is connected with the fact
that our choice of the state marking is not translational invariant.
Namely, let $\l_1=1$. Then there is such a move of the string which
increases $\alpha$ on the unity:

\be
\label{eqleft}
\mid \alpha ;1,\l_2,...,\l_m>\quad \Rightarrow \quad 
\mid \alpha +1;\l_2,\l_3,...,\l_m,n>.   
\ee

We have the analogous situation on the right ``edge'' of the torus
when the last parameter  $\l_m=n$. In this case we have:

\be
\label{eqright}
\mid \alpha ;\l_1,...,\l_{m-1},n>\quad \Rightarrow \quad 
\mid \alpha -1;1,\l_1,...,\l_{m-1}>.   
\ee

Now let us consider the operator $U$ which makes the shift up of the all
configuration on the one sell of our lattice. This operator has
the following properties:

1) It acts only on the index $\alpha$:
\be
U\mid \alpha ;\l>=\mid \alpha +1;\l>,
\label{U}
\ee

2) It commutes with the Hamiltonian $H$,

3) $\quad U^m = {\bf 1}$

Due to the second property it is useful to consider another basis which
can be obtained from the former one by
the discrete Fourie transformation in respect to the parameter 
$\alpha$:
\be
\mid \l_1,...,\l_m;a> \equiv \sum^{m-1}_{\alpha =0} e^{2\pi i a\a/m}
\mid \alpha ;\l_1,...,\l_m>,
\label{fourie}
\ee
where
\be
a=0,...,m-1.
\label{a}
\ee

Now instead of the equations
 ($\ref{eqleft}$ and $\ref{eqright}$) we have: 

\be
\label{eqleftnew}
\mid 1,\l_2,...,\l_m;a>\quad \Rightarrow \quad 
e^{-2\pi ia/ m} \mid \l_2,...,\l_m,n;a>
\ee

and the conjugated equation:

\be
\label{eqrightnew}
\mid \l_1,...,\l_{m-1},n;a>\quad \Rightarrow \quad 
e^{2\pi i a/ m} \mid 1,\l_1,...,\l_{m-1};a>.  
\ee

Below we shall omit the parameter $a$ from the marking of our new
basis vectors.

Now let us consider the wave functions instead of the basis
vectors. As it was done above we introduce the following
expansion
\be
\mid \Psi>=\sum_{\l}\Psi(\l)\mid \l>
\label{expan}
\ee

and consider the Schr{\"o}dinger equation 

\begin{equation}
\label{eqgennew1}
E\;\Psi (\l)\;=\;\sum^m_{i=1}
\Psi (\l+\d_i)+
\Psi (\l-\d_i).
\end{equation}





Then we can repeat the same arguments which were applied for derivation of the
determinant formula (\ref{det1}). As appeared we can use even more simple
substitution for the wave function

$$
\Psi (\l)=\sum_{\s}
A_{\s}x_{\s(1)}^{\l_1}\ldots x_{\s(m)}^{\l_m}.
$$
In some sence it is the simplest variant of the Bethe ansatz.

As it was discussed above for the case when $H$
acts on the state with some group of the coinciding parameters
$\l_j=\l_{j+1}=...=\l_k,\quad j<k$ we have only two allowed moves 
which change this group $\l$. Namely, the parameter $\l_j$ can be
decreased on unity while $\l_k$ can be increased on the unity.
As above this statement allows us to fix the coefficients $A_{\s}$ in the last
expresion for the wave function. The result is rather simple:

\begin{equation}
\label{det}
\Psi (\l_1,\l_2,...,\l_m)=const \times \mbox{Det} \left 
(\begin{array}{cccc}
x_1^{\l_1}&x_1^{\l_2+1}&\ldots &x_1^{\l_m+m-1} \\
x_2^{\l_1}&x_2^{\l_2+1}&\ldots &x_2^{\l_m+m-1} \\
\vdots &\vdots &\ddots &\vdots \\
x_m^{\l_1}&x_m^{\l_2+1}&\ldots &x_m^{\l_m+m-1}
\end{array} \right )
\end{equation}

As before the unwanted moves are suppressed automaticaly.

Again we have the determinant solution which in some sence corresponds
to the free fermion model.

Only one thing we have to do is to ``quantize'' the variables
$x_i$ with the help of the residual equations
(\ref{eqleftnew}) and  (\ref{eqrightnew}).

Substituting the expansion (\ref{expan}) to the conditions (\ref{eqleftnew})
and (\ref{eqrightnew}) we get the following equations for wave functions:

$$
\begin{array}{c}
\Psi (0,\l_2,...,\l_m)=e^{2\pi ia/m}\Psi (\l_2,...,\l_m,n), \\
\Psi (\l_1,...,\l_{m-1},n+1)=e^{-2\pi ia/m}\Psi (1,\l_1,...,\l_{m-1}).
\end{array}
$$

These equations are formal in the sence that we do not determine what
the wave functions 
$\Psi(0,...)$ and $\Psi(...,n+1)$ are.
If we substitute the formula
(\ref{det}) into the last pair of the relations we can obtain
from the first  one:
$$
\mbox{Det} \left 
(\begin{array}{cccc}
1&x_1^{\l_2+1}&\ldots &x_1^{\l_m+m-1} \\
1&x_2^{\l_2+1}&\ldots &x_2^{\l_m+m-1} \\
\vdots &\vdots &\ddots &\vdots \\
1&x_m^{\l_2+1}&\ldots &x_m^{\l_m+m-1}
\end{array} \right )=\eta ^a
\mbox{Det} \left 
(\begin{array}{cccc}
x_1^{\l_2}&\ldots &x_1^{\l_m+m-2}&x_1^{n+m-1} \\
x_2^{\l_2}&\ldots &x_2^{\l_m+m-2}&x_1^{n+m-1} \\
\vdots &\ddots &\vdots &\vdots \\
x_m^{\l_2}&\ldots &x_m^{\l_m+m-2}&x_1^{n+m-1}
\end{array} \right )
$$

and from the second one:

$$
\mbox{Det} \left 
(\begin{array}{cccc}
x_1^{\l_1}&\ldots &x_1^{\l_{m-1}+m-2}&x_1^{n+m} \\
x_2^{\l_1}&\ldots &x_2^{\l_{m-1}+m-2}&x_1^{n+m} \\
\vdots &\ddots &\vdots &\vdots \\
x_m^{\l_1}&\ldots &x_m^{\l_{m-1}+m-2}&x_1^{n+m}
\end{array} \right )
=\eta^{-a}\mbox{Det} \left 
(\begin{array}{cccc}
x_1&x_1^{\l_1+1}&\ldots &x_1^{\l_{m-1}+m-1} \\
x_2&x_2^{\l_1+1}&\ldots &x_2^{\l_{m-1}+m-1} \\
\vdots &\vdots &\ddots &\vdots \\
x_m&x_m^{\l_1+1}&\ldots &x_m^{\l_{m-1}+m-1}
\end{array} \right )
$$

For the sake of simplicity we have introduced the following
notation:
$$
\eta = e^{2\pi i/m}
$$

Let us consider the first equality. After some algebra we get:

\begin{equation}
\label{leq}
x_i^{n+m}=(-1)^{m+1}\eta ^{-a}\prod_{j=1}^m x_j,\quad i=1,2,...,m.
\end{equation}
It can be easily seen that the second equality gives the same condition.
Then we obtain that the relation
$x_k/x_i$ is some root of unity of the power $m+n$.
Let us introduce the following notation:
\be
\omega=e^{\frac{2\pi i}{m+n}}
\label{om}
\ee
We can look for the solution to the equation (\ref{leq})  
in the following form:
\be
x_i\;=\;\r(n,m)\;\omega^{k_i},\quad i=1,2,\ldots,m.
\label{xi}
\ee
Here as in the formula (\ref{k1})
$k_i$ are the wave numbers which satisfy
the following chain of the inequalities:

\be
0 < k_1 < k_2 < ... < k_m < m+n.
\label{k2}
\ee

Substituting it into the formula (\ref{leq}) we easily get
\be
\r(n,m)\;=\;e^{2\pi i\phi(n,m)},\quad 
\phi(n,m)\;=\;{1\over{n(n+m)}}\sum_{i=1}^m\,k_i-{a\over{nm}}-{{m+1}\over{2n}}.
\label{rho}
\ee

We have conjectured that there are no coinciding variables
$x_i$. We should note that in principle we could also consider
the coinciding $x_i$ by taking  the accurate limit. 
But the calculation of number of states shows us that
it is not necessary to do.
Indeed, the chain of inequalities (\ref{k2}) coincides with (\ref{lambda}).
Hence, after taking into account that $a=0,\ldots,m-1$ 
we come to the formula (\ref{dim2}) for the number
of all eigenvectors. 

Let us adduce the final result for the eigenvectors 
of the Hamiltonian	$H$ in case of the string of type (1,1)
\be
\Psi(\l)\;=\;\r(n,m)^{\l_1+\ldots +\l_m}\mbox{Det}(\om^{k_i(\l_j+j-1)})_{1\leq i,j\leq m}
\label{Psi11}
\ee
The spectrum of the Hamiltonian is given by
\be
E(k_1,\ldots k_m)\;=\;\sum_{i=1}^m\,2\cos{2\pi\,(\l(n,m)+{{k_i}\over{m+n}})}.
\label{E3}
\ee
We can consider the shift operator in horizonal direction $P$ (or momentum 
operator) which acts on some quantum state as follows
\be
P\mid \l_1,\ldots,\l_m>\;=\;\mid\l_1+1,\ldots,\l_m+1>\qquad P^n\;=\;I.
\label{P0}
\ee
Actually, as the operator $U$ defined by formula (\ref{U}) momentum operator
$P$ commutes with the Hamiltonian. Hence, the eigenvalue of the operator $P$
$p=\r(n,m)^m\,e^{{2\pi i b}\over n}$ where $b=0,1,\ldots n-1$ is 
the ``quantum number''.
One can see from the solution (\ref{Psi11}) that 
\be
b=\sum_{i=1}^m k_i\quad mod \quad n.
\label{be}
\ee
So, all eigenvalues and eigenvectors belong to the different $m n$ sectors
labeled by two integers $a=0,1...,m-1 $ and $b=0,1,...,n-1$.

\subsection{The string of type (1,2)}

We have considered above the spectrum problem for the string with fixed
ends and the string of type (1,1) on the torus. The determinant formulae 
(\ref{det1}) and (\ref{det}) for the wave functions have a features of
the fermionic free system. In some sense it corresponds to the non-interacting
case. Now we are going to consider the case of the string of type (1,2) 
which is already the example of the interacting string or more exactly
self-interacting string. This string winds round the torus one time in
horizontal direction and twice in the vertical one.
The example of the state of such a string is shown in Fig.~15

We can try to reduce the case of the (1,2)-string to the previous
case of the (1,1)-string by applying some simple trick.
Namely, let us consider some arbitrary configuration of (1,2)-string
on the lattice $n\times m$, for example, shown in Fig.~15.
Let us also consider the (1,1)-string configuration on the
lattice $n\times 2m$ which consists of two sheets $n\times m$.

\setlength{\unitlength}{0.3mm}

\begin{picture}(400,200)
\multiput(50,80)(20,0){13}{\line(0,1){140}}
\multiput(50,80)(0,20){8}{\line(1,0){240}}
\thicklines
\multiput(50,140)(0,1){2}{\line(1,0){40}}
\multiput(90,140)(1,0){2}{\line(0,1){20}}
\multiput(90,160)(0,1){2}{\line(1,0){20}}
\multiput(110,160)(1,0){2}{\line(0,1){40}}
\multiput(110,200)(0,1){2}{\line(1,0){20}}
\multiput(130,200)(1,0){2}{\line(0,1){20}}
\multiput(130,80)(1,0){2}{\line(0,1){20}}
\multiput(130,100)(0,1){2}{\line(1,0){40}}
\multiput(170,100)(1,0){2}{\line(0,1){20}}
\multiput(170,120)(0,1){2}{\line(1,0){20}}
\multiput(190,120)(1,0){2}{\line(0,1){40}}
\multiput(190,160)(0,1){2}{\line(1,0){20}}
\multiput(210,160)(1,0){2}{\line(0,1){20}}
\multiput(210,180)(0,1){2}{\line(1,0){40}}
\multiput(250,180)(1,0){2}{\line(0,1){40}}
\multiput(250,80)(1,0){2}{\line(0,1){40}}
\multiput(250,120)(0,1){2}{\line(1,0){20}}
\multiput(270,120)(1,0){2}{\line(0,1){20}}
\multiput(270,140)(0,1){2}{\line(1,0){20}}
\put(125,60){A}
\put(245,60){B}
\put(90,40){Fig.15 The string of type (1,2)}
\end{picture}


This configuration can be obtained from the (1,2)-string
configuration in the following way. We leave the two parts
of the string before the point A and after the point B in Fig.~15 on the
first sheet. 
The part of the string between these points A and B is
transfered to the second sheet. The result of this procedure is shown
in Fig.~16. It easy to see that the toroidal boundary conditions 
in the case of the (1,2)-string on the $n\times m$ lattice can be
naturally fulfiled for it's "double" on the $n\times 2m$ lattice. 
Namely, let us return to the original notation for the quantum state
(\ref{stringstate}) of the (1,1)-string on the $n\times 2m$ lattice.
The corresponding wave function is $\Psi(\a;\l)$ where
$\l$ denotes the quantum state corresponding to the partition
$(\l_1,\ldots,\l_{2m}),\quad 1 \leq \l_1 \leq ... \leq \l_{2m} \leq n$.
In order to recover the initial toroidal boundary conditions for
(1,2)-string we should identify two sheets discussed above. It can
be easily done by satisfying the requirement $\Psi(\a;\l)=\Psi(\a+m ;\l)$.

Applying now the Fourie transform to the wave function as in (\ref{fourie}) we obtain
\be
\Psi( \l ;a) \equiv \sum^{2m-1}_{\alpha =0} e^{2\pi i a\a/{2m}}
\Psi(\alpha ;\l)\,=\,2\sum_{\a=0}^{m-1}e^{2\pi i a\a/{2m}}\cos{{\pi a}\over 2}
\Psi(\alpha ;\l)
\label{fourie1}
\ee
where $a=0,...,2m-1$.
From (\ref{fourie1}) we see that the wave function in the LHS is not zero
if $a$ is even. Therefore we can substitute $a\rightarrow 2a$. 
The new
parameter $a=0,...,m-1$. 

Let us use again more simple notation $\Psi(\l)$ for
the wave function $\Psi(\l ;a)$ implying the dependence on this new parameter $a$.

\setlength{\unitlength}{0.3mm}
\begin{picture}(400,300)
\multiput(50,80)(20,0){13}{\line(0,1){280}}
\multiput(50,80)(0,20){15}{\line(1,0){240}}
\thicklines
\multiput(50,140)(0,1){2}{\line(1,0){40}}
\multiput(90,140)(1,0){2}{\line(0,1){20}}
\multiput(90,160)(0,1){2}{\line(1,0){20}}
\multiput(110,160)(1,0){2}{\line(0,1){40}}
\multiput(110,200)(0,1){2}{\line(1,0){20}}
\multiput(130,200)(1,0){2}{\line(0,1){20}}
\multiput(130,220)(1,0){2}{\line(0,1){20}}
\multiput(130,240)(0,1){2}{\line(1,0){40}}
\multiput(170,240)(1,0){2}{\line(0,1){20}}
\multiput(170,260)(0,1){2}{\line(1,0){20}}
\multiput(190,260)(1,0){2}{\line(0,1){40}}
\multiput(190,300)(0,1){2}{\line(1,0){20}}
\multiput(210,300)(1,0){2}{\line(0,1){20}}
\multiput(210,320)(0,1){2}{\line(1,0){40}}
\multiput(250,320)(1,0){2}{\line(0,1){40}}
\multiput(250,80)(1,0){2}{\line(0,1){40}}
\multiput(250,120)(0,1){2}{\line(1,0){20}}
\multiput(270,120)(1,0){2}{\line(0,1){20}}
\multiput(270,140)(0,1){2}{\line(1,0){20}}
\put(125,60){A}
\put(245,60){B}
\put(0,40){Fig.16 The (1,1)-string corresponding to the initial string of type (1,2)}
\end{picture}

Nevertheless the quantum problems for the (1,2)-string on the 
$n\times m$ lattice and the (1,1)-string on the $n\times 2m$ lattice
are still not comletely equivalent. To make them equivalent we should introduce
an additional restriction on the (1,1)-string which comes from a simple
observation that the length of each ``jump'' of the (1,1)-string can not 
be greater than $m$. Otherwise we would be forced to lie one part of the
string on another when making the back procedure of the comparison of
the (1,1)-string to the (1,2)-string. So, we come to the following
exclusion rule for the wave function $\Psi(\l)$ of the (1,1)-string
\be
\Psi(\l)\;=\;0\quad\mbox{if}\quad \l_i=\l_{i+1}=\ldots=\l_{i+m}
\quad\mbox{for at least one index $i$}.
\label{zapret}
\ee
We claim that if this restriction for the (1,1)-string 
is fulfiled then the quantum problems for the (1,1)-string living
on the lattice $n\times 2m$ and for the (1,2)-string living on
the lattice $n\times m$ are equivalent to each other.

In Appendix we consider the case of the (1,2)-string
for $m=2$ in more detailes. Below we shall adduce only the final
result for the general case of $m$.
But first of all let us introduce some useful notation.
Let $D$ be the generalization of the formula  (\ref{D1}) from the
Appendix

\newpage

\be
D(n_1,\ldots,n_{2m-1}|k_1,\ldots,k_{2m})\,=\,
\mbox{Det}\left 
(\begin{array}{cccc}
1&\om^{n_1k_1}&\om^{n_2k_1}&\ldots\om^{n_{2m-1} k_1}\\
1&\om^{n_1k_2}&\om^{n_2k_1}&\ldots\om^{n_{2m-1} k_2}\\
\vdots &\vdots &\ddots &\vdots\\
1&\om^{n_1k_{2m}}&\om^{n_2k_{2m}}&\ldots\om^{n_{2m-1} k_{2m}}\\
\end{array} \right )
\label{D}
\ee
where $\om=e^{{2\pi i}\over{n+2m}}$. 
Let us also define ``partial'' $D$-functions:
$$
D_{i_1,\ldots,i_{m-1}}(k_1,\ldots,k_{2m})\,=\,
D(1,2,\ldots,m,i_1,i_2,\ldots,i_{m-1}|k_1,k_2,\ldots,k_{2m})
$$
\be
D^{i_1,\ldots,i_{m-1}}(k_1,\ldots,k_{2m})\,=\,D_{i_1,\ldots,i_{m-1}}(-k_1,\ldots,-k_{2m})
\label{TD}
\ee
where $m+1<i_1<\ldots<i_{m-1}<n+2m-1$.
Let us also adduce the generalization of the matrix $A$ given by formula (\ref{A})
$$
A_{i_1,\ldots,i_{m-1}}^{j_1,\ldots,j_{m-1}}(E)\;=
$$
\be
\sum_{0\leq k_1<k_2<\ldots k_{2m-1}<2m+n}
{{D_{i_1,\ldots,i_{m-1}}(k_1,\ldots,k_{2m})D^{j_1,\ldots,j_{m-1}}(k_1,\ldots,k_{2m})}
\over{E-E(k_1,\ldots,k_{2m})}}
\label{Aform}
\ee
where $k_{2m}=b-\sum_{i=1}^{2m-1}k_i$ and $b=-n+1+m(2m-3),\ldots,m(2m-3)$
\beq
&E(k_1,\ldots,k_{2m})\;=\;\r\sum_{j=1}^{2m}\om^{k_j}\,+\,\r^{-1}\sum_{j=1}^{2m}\om^{-k_j},&
\nonumber\\
&\r=e^{2\pi i\phi},\quad\phi={b\over{n(n+2m)}} -
{a\over{nm}}-{{2m-3}\over{2n}},&\nonumber\\
&a=0,1,\ldots,m-1&.
\label{EE0}
\eeq
Let us note that $E(k_1,\ldots,k_{2m})$ is nothing else but the energy
of the Hamiltonian for (1,1)-string \footnote{
Here we have used another form 
of the solution to the (1,1)-string problem which is equivalent
to that given by formulae (4.33-4.37)
up to enumeration of  wave numbers $k_i$ and the parameter $b$.} 
Two integers $a$ and $b$ 
connected with the momentums in vertical and horizontal directions
respectively are considered to be fixed.
It is easy to see that the matrix $A$ has a dimension ${{n+m-3}\choose {m-1}}$.
The spectrum of the Hamiltonian for the case of the (1,2)-string
is determined as a solution to the ``secular'' equation
\be
\mbox{Det}A(E)\;=\;0.
\label{secular}
\ee
If we have succeeded in finding some it's solution, say $E^*$,
then there exists a zero vector $\zeta_{i_1,\ldots,i_{m-1}}$ such that
$A\zeta=0$ and the wave function for the Hamiltonian have the
following form:
\beq
&\Psi(0,n_1,\ldots,n_{2m-1})\;=\;\r^{n_1+\ldots+n_{2m-1}}\,
\sum_{m+1<j_1<\ldots<j_{m-1}<n+2m-1}\zeta_{j_1,\ldots,j_{m-1}}&\nonumber\\
&\sum_{0\leq k_1<k_2<\ldots k_{2m-1}<2m+n}
{{D(n_1,\ldots,n_{2m-1}|k_1,\ldots,k_{2m})D^{j_1,\ldots,j_{m-1}}(k_1,\ldots,k_{2m})}
\over
{E-E(k_1,\ldots,k_{2m})}}&,\nonumber\\
&\quad&
\label{Psi12form}
\eeq
where as above $k_{2m}=b-\sum_{i=1}^{2m-1}k_i$ and $b=-n+1+m(2m-3),\ldots,m(2m-3)$.
In formula (\ref{Psi12form}) we use alternative
way to determine the wave function which will be described in Appendix. Namely, 
the wave function depends on the differences $n_i=\ov n_i-\ov n_0$
and ${\ov n}_{i-1}\,=\,\l_i+i-1$.
The wave function for other combinations can be easily obtained by shifting
the coordinates on some definite number. For example, $\Psi(r,n_1+r,\ldots,n_{2m-1}+r)=
p^r\Psi(0,n_1,\ldots,n_{2m-1})$ where $p=\r^{2m}\om^b$ is an eigenvalue of the
momentum operator $P$ (see formula (\ref{P0})).

As it was mentioned above the derivation of these formulae for $m=2$ will be
given in Appendix. But it is not very difficult to understand how they work
by the direct checking that the formula (\ref{Psi12form}) with the 
equation (\ref{secular}) really gives us the solution to the eigenvalue 
problem. We have called (\ref{secular}) ``secular'' equation because it has
a similar form to the secular equation in quantum mechanics. But
in comparison with the ordinary secular equation which provides the 
second order correction to the energy our ``secular'' equation being solved 
provides the rigorous result for the energy of the string of type (1,2). 
Of course, this is a polynomial equation and we are not able to solve
it manifestly. In comparison with the ``free'' case of the (1,1)-string this
case contains some kind of the ``diffraction''. Therefore, it seems to be
non-integrable. 

\section{Discussion}

In this paper we have concentrated on some subspace of the whole 
quantum space i.e. the space of the ``string'' states. 
First of all, we have considered some more simple cases of the
string with fixed ends and the string of type (1,1) living on the
torus. Both of these cases appeared to be equivalent to some free
fermionic system. We use the wave functions of this free system
as a basis functions for the expansion in the case of the string
of type (1,2) which is the example of the ``self-interacting'' string.
The energy spectrum is given by the ``secular'' equation.
We hope that the similar result can also be obtained for other
types of the single string which has an arbitrary winding numbers
round the torus in horizontal and vertical directions. 
We think that the case of two and more string is not more complicated
in comparison with the case of the single string. 
It could be also reasonable to consider the ``scattering'' of two and more
strings with the fixed ends. Some preliminary analysis shows that
the picture of the interaction in case of two strings with the fixed
ends is very interesting and looks rather non-trivial.

We also hope that the investigation of the thermodynamic limit can be
done for the interacting case as well. Perhaps, it will demand the
introduction of some other parameters because the Hamiltonian considered
here corresdponds to the simplest variant and can be, in priciple,
generalized by introducing new terms with some arbitrary coefficients
as in XXZ or XYZ spin chains.

\section{Acknowledgements}

The author would like to thank Yu.G.~Stroganov who informed author about
the "string" model considered in this paper. 
We would also like to thank R.~Flume, V.V.~Mangazeev, G.P.~Pronko and S.M.~Sergeev
for stimulating discussions and suggestions.
The author is also grateful to
R.~Flume for his kind hospitality in the Physical Institute of
Bonn University.
This research has been  supported 
by Alexander~von~Humboldt Foundation. 

\noindent

\section{Appendix}
Here we consider the case $m=2$ of the string of type (1,2). 
As we claimed above we have to consider the equivalent problem
of the (1,1)-string	 on the lattice $n\times 4$ with the
restriction that not greater than two neightbouring ``coordinates''
of jumps $\l_i$ can coincide with each other (see formula (\ref{zapret})).
Let $\mid \l_1,\l_2,\l_3,\l_4; a>$ where $a=0,1$ and 
$1\leq\l_1\leq\l_2\leq\l_3\leq\l_4\leq n$ 
is the quantum state obtained by the Fourie transformation as in
formula (\ref{fourie1}). 
It is more convenient to mark these states in terms of the variables
${\ov n}_{i-1}\,=\,\l_i+i-1$ which satisfy 
\be
0<{\ov n}_0<{\ov n}_1<{\ov n}_2<{\ov n}_3<N,
\label{neqforn}
\ee
where $N=n+4$.
Namely, droping again the dependence on $a$ we get now notation for the quantum state
$\mid {\ov n}_0,{\ov n}_1,{\ov n}_2,{\ov n}_3>$. 
Let $P$ be the shift operator in the horizontal direction
\be
P\mid {\ov n}_0,{\ov n}_1,{\ov n}_2,{\ov n}_3>\,=\,
\mid {\ov n}_0+1,{\ov n}_1+1,{\ov n}_2+1,{\ov n}_3+1>
\label{P}
\ee
and $P^n=I$.
Let us also define the state $\mid 0,n_1,n_2,n_3>$ as follows
\be
\mid 0,{\ov n}_1-{\ov n}_0,{\ov n}_2-{\ov n}_0,{\ov n}_3-{\ov n}_0>\;=
\;P^{-{\ov n}_0}\mid {\ov n}_0,{\ov n}_1,{\ov n}_2,{\ov n}_3>.
\label{state0}
\ee
So, using this definition we can deel only with states of the form $\mid 0,n_1,n_2,n_3>$
where $n_1,n_2,n_3$ are used instead of the differences 
${\ov n}_1-{\ov n}_0,{\ov n}_2-{\ov n}_0,{\ov n}_3-{\ov n}_0$ in the LHS of (\ref{state0}). 
If $n_1>1$ then doing one step down we can easily get
\be
\mid 0,n_1,n_2,n_3>\;=\;(-1)^a\, \mid n_1-1,n_2-1,n_3-1,N-1>.
\label{stepdown1}
\ee
If $n_1=1$ we can do two steps down. Then
\be
\mid 0,1,n_2,n_3>\;=\; \mid n_2-2,n_3-2,N-1,N>.
\label{stepdown2}
\ee

Shifting consequently the state $\mid 0,n_1,n_2,n_3>$ on $n_1-1$,$n_2-2$ and $n_3-3$
steps in the horizontal direction and using (\ref{P}) and (\ref{stepdown1},\ref{stepdown2}) 
we obtain the following chain of the equalities
\beq
&\mid 0,n_1,n_2,n_3>\;=\;(-1)^aP^{n_1-1}\mid 0,n_2-n_1,n_3-n_1,N-n_1>\;=\;&\nonumber\\
&P^{n_2-2}\mid 0,n_3-n_2,N-n_2,N+n_1-n_2>\;=\;&\nonumber\\
&(-1)^a P^{n_3-3}\mid 0,N-n_3,N+n_1-n_3,N+n_2-n_3>&.
\label{chainstate0}
\eeq

Let us call the states 
for which any three variables $n_i,n_{i+1},n_{i+2}$ go in successive way
(for example $0,1,2$) ``forbidden'' states. In fact, we have only 
four possibilities for the forbidden states. Namely, the states of a types\\
$\mid0,1,2,n_3>$,$\mid 0,n_1,n_1+1,n_1+2>$, $\mid 0,n_1,N-2,N-1>$ and\\
$\mid 0,1,n_2,N-1>$ are forbidden.
The wave function must be zero on such a states 
in accordance with (\ref{zapret}).

Let $H_0$ be the Hamiltonian 
for the (1,1)-string which does not ``distinguish'' the ``forbidden'' states from
other states satisfying only the requirement (\ref{neqforn}).
The Hamiltonian $H$ for which we want to solve the eigenvalue problem is
\be
H\;=\;H_0\,+\,\d H
\label{Hint}
\ee
where the interaction $\d H$ acts non-trivially only on the ``forbidden'' states.
Actually, the action of $\d H$ should be so that the result of it's action
on some forbidden state would compensate a result of the action of $H_0$
on this state. We should note that the resulting states
can not be forbidden already. We have to take into consideration
only such a states. The result of action of $\d H$ on the arbitrary state
$\mid 0,n_1,n_2,n_3>$ is a sum of four terms in accordance with four possiblities
to get the forbidden states mentioned above. Namely, we have
\beq
&\d H\, \mid 0,n_1,n_2,n_3>\;=\;
-\d_{n_1,1}\d_{n_2,2}(P^{-1} \mid 0,2,3,n_3+1>+ \mid 0,1,3,n_3>)-&\nonumber\\
&\d_{n_2,n_1+1}\d_{n_3,n_1+2}(-1)^a P^{n_1-1}
(P^{-1} \mid 0,2,3,N-n_1+1>+ \mid 0,1,3,N-n_1>)&\nonumber\\
&-\d_{n_2,N-2}\d_{n_3,N-1}(P^{-1} \mid 0,2,3,n_1+1>+ \mid 0,1,3,n_1+2>)&\nonumber\\
&-\d_{n_1,1}\d_{n_3,N-1}(-1)^a(P^{-1} \mid 0,2,3,n_2+2>+ \mid 0,1,3,n_2+1>)&.
\label{dH}
\eeq

As in Section 4 we are going to solve the eigenvalue problem
\be
(H_0+\d H)\,\mid \Psi>\;=\;E\,\mid\Psi>
\label{Heig}
\ee
with the help of the expansion
\be
\mid\Psi>\;=\;\sum_{0<{\ov n}_0<{\ov n}_1<{\ov n}_2<{\ov n}_3<N}
\Psi({\ov n}_0,{\ov n}_1,{\ov n}_2,{\ov n}_3)\mid {\ov n}_0,{\ov n}_1,{\ov n}_2,{\ov n}_3>,
\label{exp}
\ee
where the contribution from the forbidden states are suppressed by the
requirement that the wave function $\Psi({\ov n}_0,{\ov n}_1,{\ov n}_2,{\ov n}_3)$ 
is zero for them.

Now one can apply the technique which is rather standard in quantum mechanics. 
Namely, we can look for the wave function as an expantion
on the basis of the eigenfunctions of the operator $H_0$ and then
use the completness of these eigenfunctions.
So, using the result (\ref{Psi11}) with the different enumeration of the
wave numbers $k_i$ and parameter $b$ we get

\newpage

$$
\Psi(0,n_1,n_2,n_3)\;=\;
\sum_{0\leq k_1<k_2<k_3<N}\,\r^{n_1+n_2+n_3}C(k_1,k_2,k_3)\,
$$
\be
D(n_1,n_2,n_3|k_1,k_2,k_3,k_4),
\quad\label{subst}
\ee
where $k_4=b-k_1-k_2-k_3$ and $b=-n+3,\ldots,2$ is supposed to be fixed  
\be
D(n_1,n_2,n_3|k_1,k_2,k_3,k_4)=\mbox{Det}
\left 
(\begin{array}{cccc}
1&\om^{n_1k_1}&\om^{n_2k_1}&\om^{n_3k_1}\\
1&\om^{n_1k_2}&\om^{n_2k_2}&\om^{n_3k_2}\\
1&\om^{n_1k_3}&\om^{n_2k_3}&\om^{n_3k_3}\\
1&\om^{n_1k_4}&\om^{n_2k_4}&\om^{n_3k_4}
\end{array} \right )
\label{D1}
\ee 
\be
\r\;=\;e^{2\pi i\phi},\quad \phi\;=\;{1\over{nN}}b - {{a+1}\over{2n}},
\quad a=0,1
\label{r1}
\ee
and $\om=e^{{2\pi i}\over N}$.
Substituting (\ref{subst}) into the equation (\ref{Heig}) and taking
into account that the eigenvalues of the Hamiltonian $H_0$ are
\be
E(k_1,k_2,k_3,k_4)\;=\;\r\sum_{i=1}^4\om^{k_i}\,+\,\r^{-1}\sum_{i=1}^4\om^{-k_i}
\label{EofH0}
\ee
we get the equation for the unknown coefficients $C(k_1,k_2,k_3)$
\beq
&\r^{n_1+n_2+n_3}\sum_{0\leq k_1<k_2<k_3<N}&\nonumber\\
&(E-E(k_1,k_2,k_3,k_4))C(k_1,k_2,k_3)
D(n_1,n_2,n_3|k_1,k_2,k_3,k_4)\;=\;&\nonumber\\
&\quad&\nonumber\\
&\d_{n_1,1}\d_{n_2,2}S(n_3)+\d_{n_2,n_1+1}\d_{n_3,n_1+2}(-1)^a p^{n_1-1}S(N-n_1)&
\nonumber\\
&+\d_{n_2,N-2}\d_{n_3,N-1}S(n_1+2)+\d_{n_1,1}\d_{n_3,N-1}S(n_2+1)&
\label{urnaC}
\eeq
where $S(j)\,=\,-p^{-1}\Psi(0,2,3,j+1)-\Psi(0,1,3,j)$ and 
$p=\r^4\om^b$ is eigenvalue of the shift operator $P$ defined
by (\ref{P}).
Now we can use the fact that the wave function $\Psi$ is zero for
the forbidden configurations. So, we come to $S(3)\,=\,S(N-1)\,=\,0$.

Let us multiply the LHS and RHS of the equation (\ref{urnaC}) on 
$$
\r^{-n_1-n_2-n_3}D(n_1,n_2,n_3|-l_1,-l_2,-l_3,-l_4)
$$ 
with $0\leq l_1<l_2<l_3<N$ and $l_4=b-l_1-l_2-l_3$ and take a sum
over $0\leq n_1<n_2<n_3\leq N-1$ using the orthohonality property
\beq
&\sum_{0\leq n_1<n_2<n_3\leq N-1}D(n_1,n_2,n_3|-l_1,-l_2,-l_3,-l_4)
D(n_1,n_2,n_3|k_1,k_2,k_3,k_4)\;=\;&\nonumber\\
&\kappa \d_{k_1,l_1}\d_{k_2,l_2}\d_{k_3,l_3}&,
\label{orth}
\eeq
where $\kappa$ is some constant. In this formula we imply that adding to
$k_4$ and $l_4$ the period $N$ enough many times $0\leq {\ov k}_4=k_4+s N\leq N-1$
and $0\leq {\ov l}_4=l_4+t N\leq N-1$ the new numbers $\ov k_4$ and $\ov l_4$ do
not coincide with one of $k_1,k_2,k_3$ and $l_1,l_2,l_3$ respectively and
the sequences $k_1,k_2,k_3,{\ov k}_4$ and $l_1,l_2,l_3,{\ov l}_4$ are
ordered in the same way, for example, $k_1<k_2<{\ov k}_4<k_3$ and
$l_1<l_2<{\ov l}_4<l_3$.

After this we get
\beq
&(E-E(l_1,l_2,l_3,l_4))C(l_1,l_2,l_3)\;=\;&\nonumber\\
&\quad&\nonumber\\
&\sum_{n_3=4}^{N-2}\r^{-3-n_3}D(1,2,n_3|-l_1,-l_2,-l_3,-l_4)S(n_3)&\nonumber\\
&\sum_{n_1=2}^{N-4}(-1)^a p^{n_1-1}\r^{-3-3n_1}
D(n_1,n_1+1,n_1+2|-l_1,-l_2,-l_3,-l_4)S(N-n_1)&\nonumber\\
&\sum_{n_1=2}^{N-4}\r^{-n_1-2N+3}D(n_1,N-2,N-1|-l_1,-l_2,-l_3,-l_4)S(n_1+2)&\nonumber\\
&\sum_{n_2=3}^{N-3}(-1)^a\r^{-n_2-N}D(1,n_2,N-1|-l_1,-l_2,-l_3,-l_4)S(n_2+1)&.
\label{urnaC1}
\eeq
After some algebra we come to conclusion that all four terms in the RHS of
(\ref{urnaC1}) are equal to each other. So, we get the forllowing expression for 
the coefficients $C$
\be
C(k_1,k_2,k_3)\;=\;\sum_{\mu=4}^{N-2}\zeta(\mu){{D(1,2,\mu|-k_1,-k_2,-k_3,-k_4)}
\over{E-E(k_1,k_2,k_3,k_4)}},
\label{C}
\ee
where $\zeta(\mu)\,=\,4\r^{-\mu-3}S(\mu)$.
Substituting this result for $C$ into the formula (\ref{subst}) we get the
expression for the wave function

\newpage

\beq
&\Psi(0,n_1,n_2,n_3)\,=\,\r^{n_1+n_2+n_3}\sum_{0\leq k_1<k_2<k_3<N}&\nonumber\\
&\sum_{\mu=4}^{N-2}\zeta(\mu)
{{D(n_1,n_2,n_3|k_1,k_2,k_3,k_4)D(1,2,\mu|-k_1,-k_2,-k_3,-k_4)}
\over{E-E(k_1,k_2,k_3,k_4)}}&\nonumber\\
&\quad&\label{Psi12m=2}
\eeq
where we imply again that $k_4=b-k_1-k_2-k_3$.
The last step we should do is to satisfy the requirement that $\Psi=0$ for the
forbidden configuraions. In fact, due to the cyclicity property it is enough
to satisfy only 
\be
\Psi(0,1,2,\mu)\;=\;0.
\label{zap}
\ee
So, we come to the condition
\be
\sum_{\nu=4}^{N-2}A_{\mu\nu}(E)\zeta(\nu)\;=\;0,
\label{0vector}
\ee
where $A$ has the matrix elements
\beq
&A_{\mu\nu}(E)\;=&\nonumber\\
&\sum_{0\leq k_1<k_2<k_3<N}
{{D(1,2,\mu|k_1,k_2,k_3,k_4)D(1,2,\nu|-k_1,-k_2,-k_3,-k_4)}\over{E-E(k_1,k_2,k_3,k_4)}}&
\nonumber\\
&\quad\mbox{for}\quad 4\leq\mu,\nu\leq N-2&\nonumber
\label{A}
\eeq
where as above we imply that in the sum $k_4=b-k_1-k_2-k_3$.

So, the following condition should be valid
\be
\mbox{Det}A(E)\;=\;0.
\label{detA}
\ee
This condition can be considered as the equation on the energy $E$. The solutions
to this equation give us the spectrum for the Hamiltonian $H$ for the string
of type (1,2) in the case $m=2$.


\begin{thebibliography}{**}
\bibitem{Bethe}
H.~Bethe, {\it Zeitschrift fur Physik} {\bf71} (1931) 205-226.
\bibitem{Heischain}
W.~Heisenberg, {\it Zeitschrift fur Physik} {\bf49} (9-10) (1928) 619-636.
\bibitem{Bax}
R. J. Baxter, {\it Exactly Solved Models in Statistical Mechanics},
New York: Academic Press, (1982).
\bibitem{free-ferm}
E.~Lieb, T.~Schultz and D.~Mattis, {\it Annals of Physics} {\bf 16}, (1961) 407-466;\\
C.~Fan and F.Y.~Wu, {\it Phys. Rev.} {\bf B2} (1970) 723-733.
\bibitem{Lieb}
E.~Lieb  {\it Phys. Rev.} {\bf 162} (1967) 162-172;\\
E.~Lieb{\it Phys. Rev. Lett} {\bf 18} (1967) 1046-1048;{\bf 19} (1967) 108-110.
\bibitem{BQ}
R.J. Baxter, G.R.W. Quispel, {\it Journ. Stat. Phys.} {\bf 58}
n. 3/4, (1990) 411-430.
\bibitem{Z1}
A.B. Zamolodchikov, {\it Zh. Eksp. Teor. Fiz.} {\bf 79} (1980) 641-664
[English transl.: {\it JETP} {\bf 52} (1980) 325-336]
\bibitem{Z2}
A.B. Zamolodchikov, {\it Commun. Math. Phys.} {\bf 79} (1981) 489-505.
\bibitem{Strog}
Yu.G.~Stroganov, private communication, 1995.
\bibitem{Mac}
I.G.~Macdonald, {\it Symmetric Functions and Hall Polynomials},
Oxford: Clarendon Press, (1979).
\end{thebibliography}
\end{document}